\documentclass[doublecol]{epl2} 

\usepackage{amsmath}

\title{Contributions from weak-interaction channels to inclusive hadron spectra at TeV energies}
\shorttitle{Contributions from weak-interaction channels at TeV energies} 

\author{
A. P\'asztor\inst{1,2}\footnote{apasztor@rmki.kfki.hu}, 
P. L\'evai\inst{1}\footnote{plevai@rmki.kfki.hu}, \and 
Z. Tr\'ocs\'anyi\inst{3,4}\footnote{Zoltan.Trocsanyi@cern.ch} }
\shortauthor{A. P\'asztor, P. L\'evai, Z. Tr\'ocs\'anyi}

\institute{                    
  \inst{1} MTA KFKI RMKI, Konkoly-Thege Mikl\'os \'ut 29-33, 
           H-1121 Budapest, Hungary\\
  \inst{2} Roland E\"otv\"os University, 
           H-1117 P\'azm\'any s\'et\'any 1/A, Budapest, Hungary \\
  \inst{3} University of Debrecen, H-4001 Debrecen P.O.Box 51, Hungary \\
  \inst{4} Institute of Nuclear Research 
           of the Hungarian Academy of Sciences,
           H-4010 Debrecen P.O.Box 105, Hungary
}

\abstract{At the FERMILAB TEVATRON and CERN LHC accelerators partonic
processes reach such a high energy region, where the weak interactions
could enter into consideration in the hadron yields
through the opening of resonant W and Z exchange diagrams. 
We investigate these contributions to inclusive charged hadron
production in $p+p$ and $p+ {\bar p}$ reactions at a few TeV energies,
and compare our results to existing CDF and CMS data.  We perform a
leading order perturbative QCD calculation and include electroweak
channels to estimate the order of magnitude of charged hadron excess. 
The energy dependence of hadron production and the fine structure of
transverse momentum distributions are investigated. As expected, in
$p+ {\bar p}$ collisions electroweak channels yield much larger
contribution, than in $p+p$ collisions. However, these types of
contributions remains close to be negligible at the available
accelerator energies.  }

\begin{document}

\maketitle

\section{Introduction}

The first year of successful run of CERN Large Hadron Collider resulted
enormous amount of data in $p+p$ collisions at energies 
$\sqrt{s}=  \ 2.36, \ 7 \ \rm{TeV}$. The preliminary data 
of the ALICE~\cite{ALICE}, ATLAS~\cite{ATLAS} and CMS~\cite{CMS}
collaborations on inclusive charged hadron production 
display the extension of transverse momentum spectra well measured up
to 100 GeV/c or even beyond.
These results complement results from the CDF Collaboration 
at FERMILAB TEVATRON. They
reported~\cite{CDF,CDFerr} the inclusive charged hadron distribution 
measured up to $p_T \approx 50 - 100 \ \rm{GeV/c}$ at
$\sqrt{s} = 1.96 \ \rm{TeV}$ collision energy.
With increasing luminosity and running time we can expect sufficiently 
large number of recorded events at LHC so that the inclusive
hadron spectra will be measured up to
transverse momenta of couple of hundred GeV
with  high precision.
Thus all contributions beyond standard channels based on strong
interaction particle production will be revealed, if they have
small but non-negligible effect.

Inclusive hadron production at high transverse momenta is used to be 
calculated in the framework of perturbative quantum chromodynamics (QCD)
 improved parton models, which reproduces the
well known scaling properties of the measured hadron spectra, namely
$E\frac{d^3\sigma}{d^3p} \propto p_T^{-n}$, where $n \approx 5$.
When electroweak channels from W and Z boson exchange start to give
noticeable contributions then one would expect the above scaling
to be violated because of the resonance structure of the heavy boson
exchange. This violation appears at half of the bosonic mass values
at jet levels, but shifted down to lower momentum values 
in hadronic spectra as a consequence of jet fragmentation.

In this letter we address a general question.
The motivation of our work is to quantify the role of 
electroweak channels in inclusive hadron production.
Recently, the CDF collaboration~\cite{CDF} reported a very strong 
scaling violation in inclusive charge hadron spectra at high transverse 
momentum, which triggered considerable theoretical interest.
In Ref.~\cite{Albino} the fragmentation functions have been
investigated carefully, and no means have been found to explain these CDF data.
In another effort the inclusion of electroweak channels
were proposed to explain the CDF data~\cite{Ioffe}.
Here we report the outcome of our attempt to repeat this calculation.
Although we make several approximations, we carefully investigate their
possible effect and conclude that {\em both at TEVATRON and at
LHC the yield of the electroweak channels remains below the per cent
level as compared to the yield of the strong processes}.

In the meantime, the CDF collaboration published a paper with
reanalyzed data on inclusive hadron spectra~\cite{CDFerr}, invalidating
their earlier findings.
The strong scaling violation reported before almost entirely disappeared,
although the new data stop at smaller transverse momentum value
($p_T \leq 50 \ \rm{GeV/c}$), than the firstly reported dataset
($p_T \leq 100 \ \rm{GeV/c}$).
Also recently, the CMS collaboration has published their data at
$\sqrt{s}= 7 \ \rm{TeV}$ and no strong violation was reported. 
Thus, it seems that strong scaling violations do not appear in the
inclusive hadron spectra at current colliders. Nevertheless, we believe
that for future reference it is useful to publish our quantitative
predictions with the electroweak production channels also taken into
account.

\section{Pure perturbative QCD framework}

The inclusive hadron spectra is determined in the perturbative QCD 
improved parton model~\cite{FieldQCD}:
\begin{equation}
\begin{split}
& \left(E \frac{d^3\sigma}{d^3p}\right) = 
  \sum\int_0^1dx_a\int_0^1dx_b\int_0^1dz_c \times \\
& \times f_{a/p}(x_a,Q^2)f_{b/p(\bar{p})}(x_b,Q^2)
   D_{h/c}(z_c,Q_F^2)\frac{\hat{s}}{\pi z_c^2} \times \\
& \times \left( \frac{d\sigma}{d\hat{t}}\right) 
  \delta(\hat{s}+\hat{t}+\hat{u}),
\end{split}
\end{equation}
where all reactions with $u$, $d$, $s$ or $c$ quarks in the incoming or
outgoing channels are summed over. Here $f_{a/p}$ denotes the
parton distributions function of parton $a$ in the proton,
and $D_{h/c}$ is the fragmentation function of hadron $h$ emerging from
parton $c$. For factorization scales 
we use $Q=\kappa p_{T\, {\rm jet}}=\kappa p_T / z_c$, and $Q_F=\kappa p_T$,
with $\kappa = 1$ as default.  
We use the MSTW2008 parton distribution functions~\cite{MSTW} 
and the KKP fragmentation functions~\cite{KKP} for our leading order (LO)
calculation.

\section{Estimating the weak-interaction contributions}

\begin{figure}[t]
\begin{center}
\includegraphics{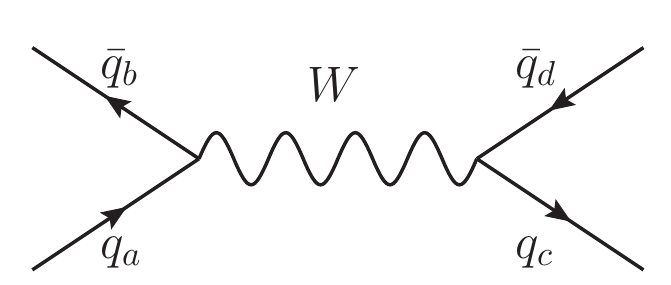}
\caption{The dominant weak-interaction mediated Feynman diagrams with a hadronic final state.}
\label{fig:feyn_diag}
\end{center}
\end{figure}

Here we estimate the order of magnitude of the weak-interaction contributions
by using a leading order calculation in the parton model. 
The dominant W boson exchange channels are 
$q_a \bar{q_b} \to W^{\pm} \to q_c\bar{q_d}$, 
as illustrated by Fig.~\ref{fig:feyn_diag}.
We neglect Z boson exchange and $t$-channel diagrams.
This is justified by the fact that the coupling of Z to quarks is
smaller than the W coupling, and t-channel diagrams are suppressed
as compared to the resonant $s$-channel contributions.
We also neglect photons, as these do not posses a resonance structure, 
therefore their contribution is also expected to drop rapidly with $p_T$.  

The Born-level cross section for the partonic subprocess shown in
Fig.~\ref{fig:feyn_diag} is
\begin{equation}
\label{equ:parton_level_xs}
\left( \frac{d\sigma}{d\hat{t}} \right)_W = \frac{g^4}{64\pi \hat{s}^2} 
|V_{ab}|^2|V_{cd}|^2
\frac{\hat{u}^2}{\left((\hat{s}-m_W^2)^2+m_W^2\Gamma_W^2\right)^2}\,,
\end{equation}
where $\hat{s}$ and $\hat{u}$ are the partonic Mandelstam variables,
$m_W$ and $\Gamma_W$ are the mass and width of the W boson, 
$V$ is the CKM matrix, and $g^2=e^2/sin^2\theta_W$ is the weak coupling.
Since the main contribution comes from $\hat{s}$ values close to the W 
mass, we can use the weak coupling measured at the vector boson mass,
neglecting the running of the electroweak coupling, which has weaker
momentum dependence, than $\alpha_s$. Thus we use $\alpha(M_Z) = 1/127.9$,
and the weak mixing angle $sin^2\theta_W(M_Z) = 0.231$ (see \cite{PDG}). 

\section{Difference between $pp$ and $p{\overline p}$ collisions}

Considering usual strong interaction production channels the hadron yields 
are very similar in $pp$ and $p{\overline p}$ collisions. The 
slight deviation in the parton distribution function for valence and sea 
quarks will generate a very small difference in the 
production of charged hadrons.

In earlier investigation of on-shell $W^+$ and $W^-$ production, an asymmetry
has been found~\cite{Szczurek2003}, which strongly depends on the
${\overline d} - {\overline u}$ asymmetry in proton distribution functions.
Similar effect appears in weak boson mediated parton-parton scattering.
We expect a larger contribution from $W^{\pm}$-channels
in $p{\overline p}$ collisions relative to the yield in $pp$ collisions
because in the ${\overline p}$ the weight of valence antiquark is
much larger than the the weight of the sea antiquarks in $p$.

\section{Numerical results}

We calculate hadron production for $p+\bar{p}$ collisions at
$\sqrt{s}=1.96 \ \rm{TeV}$,  and compare the numerical results to the
CDF data~\cite{CDF, CDFerr}.  We repeat the investigation in our pQCD
based calculation at 7 TeV in $p+p$ collisions and compare the
numerical results to CMS data.  The factorization scale dependence was
investigated by varying the $\kappa$ values between $1/2$ and $3/2$. 
Figure 2 and 3 display the obtained results. The data points are from
CMS~\cite{CMS}, and CDF~\cite{CDF, CDFerr} respectively, which are
close to the pQCD LO predictions.

\begin{figure}[t]
\begin{center}
\includegraphics[width=\linewidth]{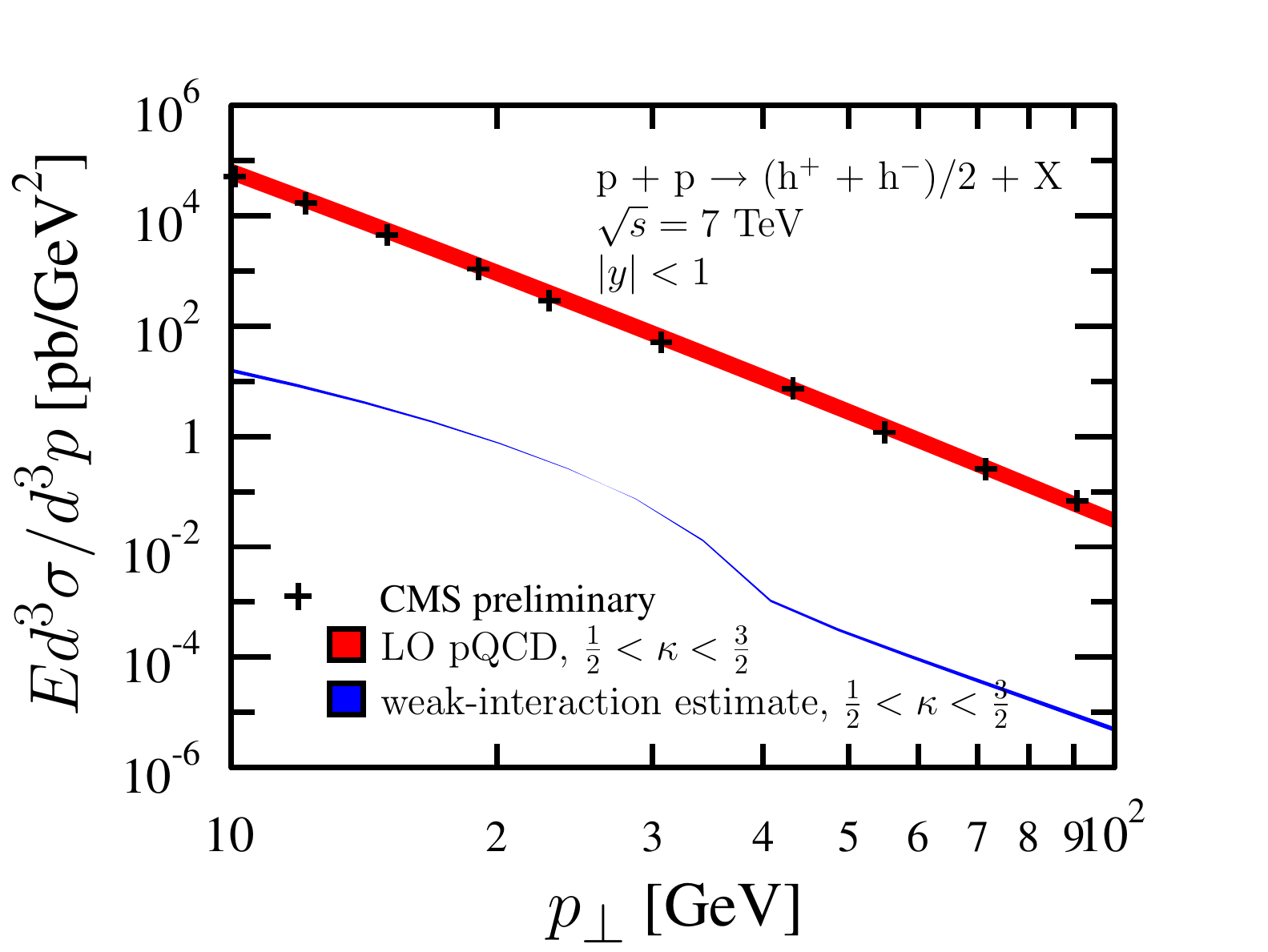}
\end{center}
\caption{({\em Color online.}) Contributions to inclusive average
charged particle yield from strong (upper red band) and weak (bottom
blue band) interaction channels in $pp$ collisions at $\sqrt{s}=7 \
\rm{TeV}$ energy.  The scale uncertainty of the strong and weak
interactions is indicated by the bands.  Experimental data are from the
CMS Collaboration~\cite{CMS}.}
\label{fig:cmscdf1}
\end{figure}
\begin{figure}[t]
\begin{center}
\includegraphics[width=\linewidth]{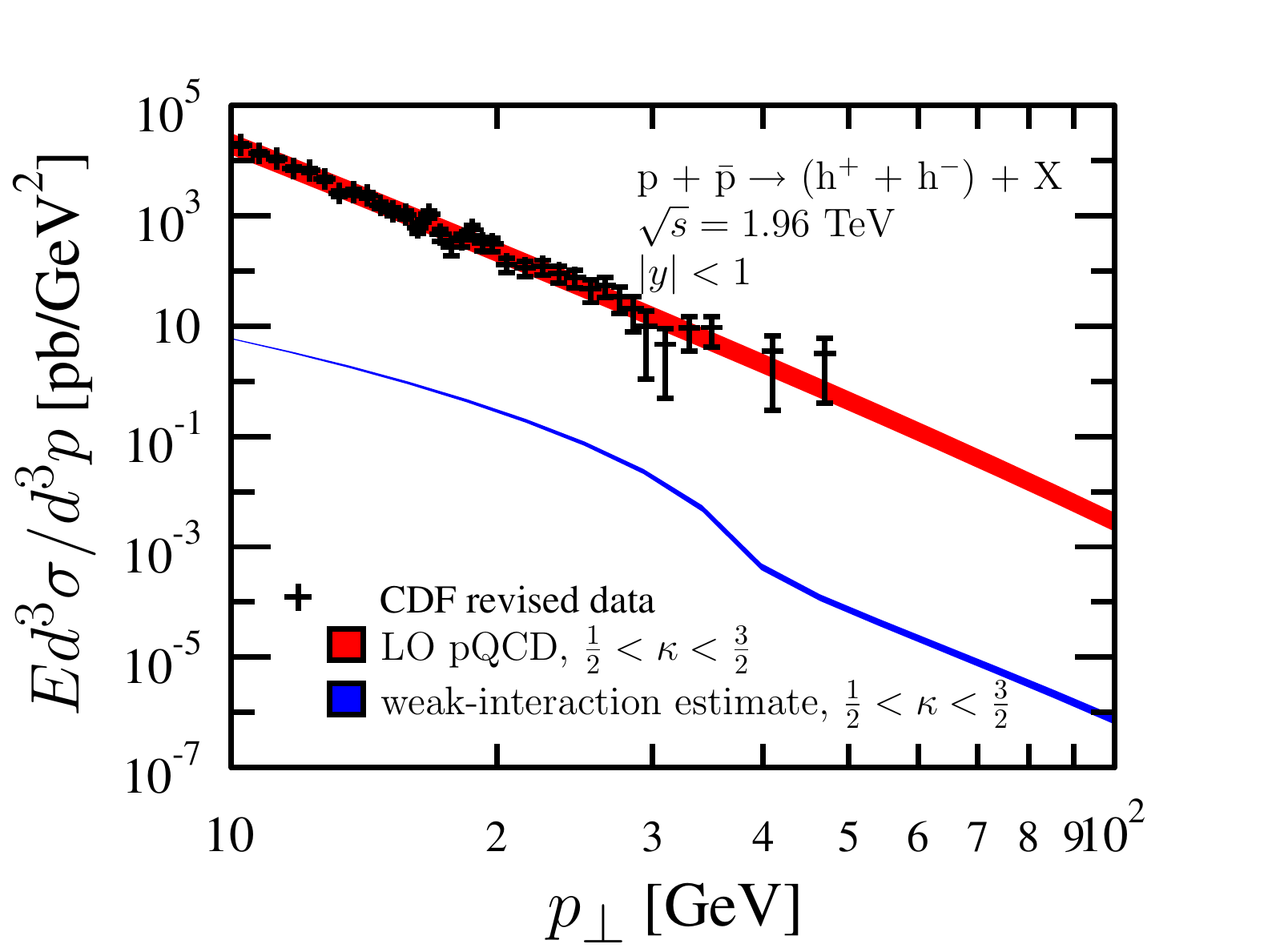}
\end{center}
\caption{({\em Color online.}) Contributions to inclusive charged
particle yield from strong (upper red band) and weak (bottom blue
band) interaction channels in $p\bar{p}$ collisions at $\sqrt{s}=1.96 \
\rm{TeV}$ energy.  The scale uncertainty of the strong and weak
interactions is indicated by the bands.  Experimental data are from the
CDF Collaboration~\cite{CDFerr}.}
\label{fig:cmscdf2}
\end{figure}

Our results indicate that weak channel contribution is more than three
orders of magnitude smaller then strong channel contributions.
The choice of the scales do not influence significantly this result.

One would expect a bump in the jet spectrum at $p_T \approx m_W / 2$
from the decay of W bosons, and a bump in the hadron spectrum at
somewhat lower momenta, due to fragmentation of quarks to lower
momentum hadrons.  Indeed our calculation of the weak-interaction
contribution reproduces this feature.

\begin{figure}[t]
\includegraphics[width=\linewidth]{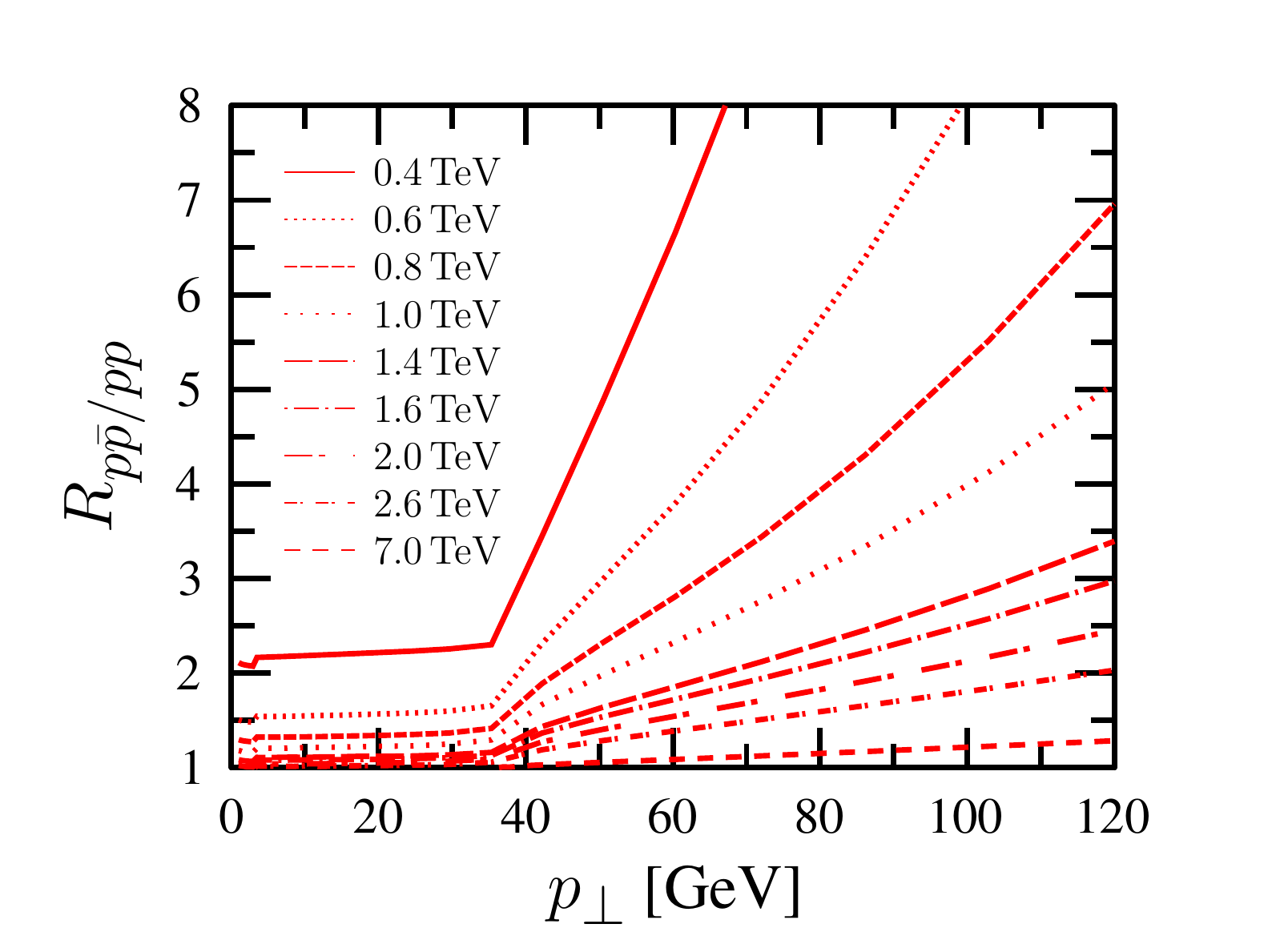}
\caption{({\em Color online.}) The ratio of weak-interaction
contributions in $p+\bar{p}$ to $p+p$ collisions at different
$\sqrt{s}$ energy values.}
\label{fig:pbarp}
\end{figure}

As mentioned earlier, the weak-interaction contribution should be
considerably higher in $p+\bar{p}$ collisions than in $p+p$. To
investigate this difference, we made calculations at different center
of mass energies. We define the ratio:
\begin{equation}
R_{p\bar{p}/pp} = 
\frac{\left(E \frac{d^3\sigma }{d^3p}\right)_W(p+\bar{p} \to h^{\pm} + X)}
{\left(E \frac{d^3\sigma }{d^3p}\right)_W(p+p \to h^{\pm} + X)},
\end{equation}
 and plot this quantity on Fig.\ref{fig:pbarp}. 
During this study, we fix $\kappa = 1$. The main difference 
in the curves for different  $\sqrt{s}$ energy values
is that in higher energy collisions the ratio 
$R_{p\bar{p}/pp}$ is smaller.
This can be readily understood by the increasingly larger weight of
the gluon in the proton with decreasing partonic momentum fraction $x$
because the higher the collision energy, the smaller $x$ contribute to
the produced hadron with the same transverse momentum.
The ratio increases with increasing $p_T$ 
due to the increasing valence quark contribution, coming from the higher
momentum fraction part of the parton distribution functions. 

We also define the ratio of the weak and strong interaction 
contributions in $p\bar{p}$ collisions:
\begin{equation}
R_{W/S} = \frac{\left(E \frac{d^3\sigma }{d^3p}\right)_W(p+\bar{p} \to h^{\pm} + X)}
{\left(E \frac{d^3\sigma }{d^3p}\right)_S(p+\bar{p} \to h^{\pm} + X)}.
\end{equation}

We investigate the dependence of this ratio on the collision
energy and the transverse momentum. Our result are displayed on 
Fig. 5.  We also used $\kappa = 1$ here.  As expected, the ratios have
a peak at somewhat lower transverse momenta than $M_W/2$. The ratio is
higher in lower energy collisions because of the dominance of gluons at
small $x$, discussed above.
\begin{figure}[t]
\includegraphics[width=\linewidth]{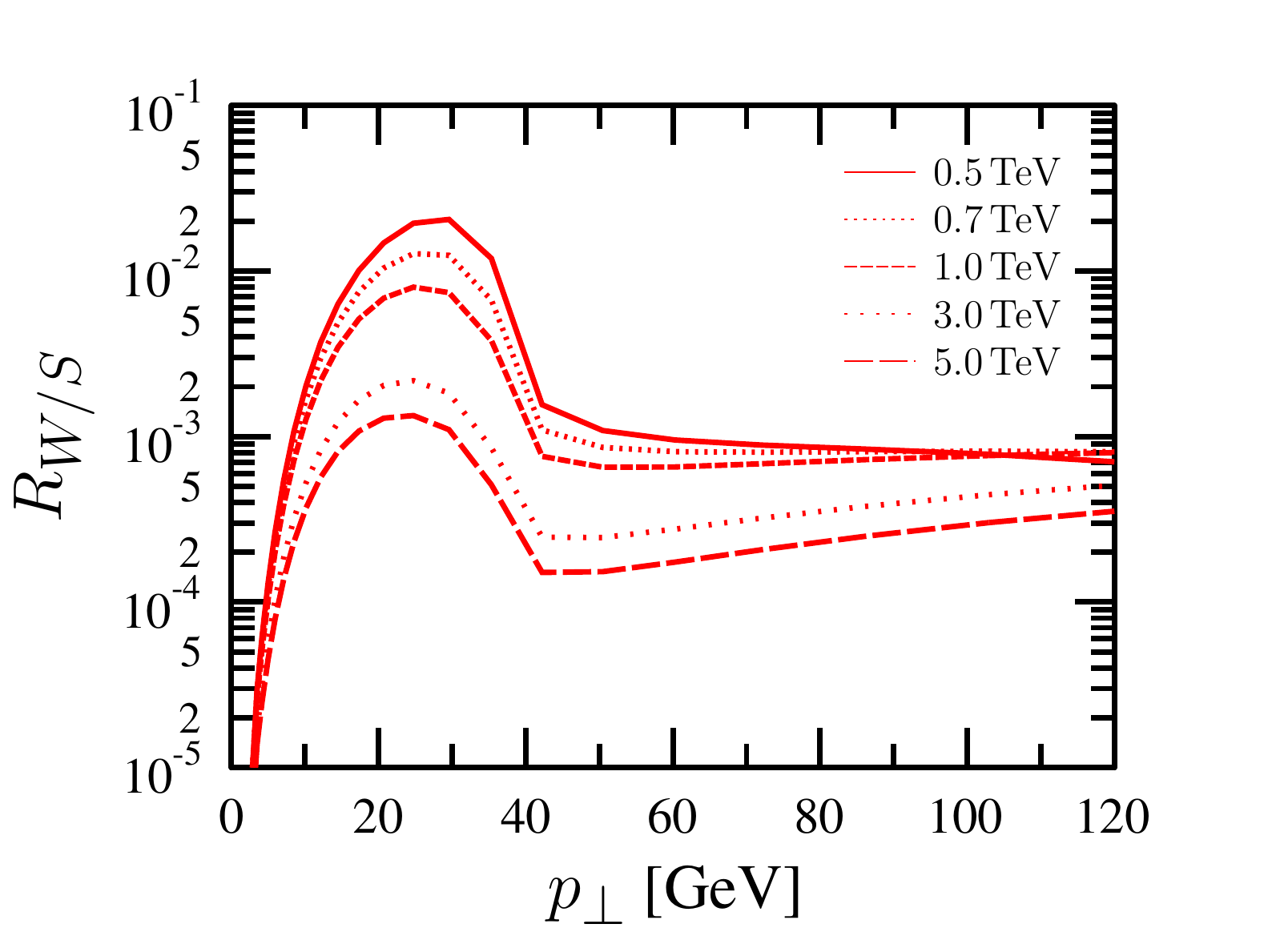}
\caption{({\em Color online.})
The ratio of weak-interaction contributions to QCD contributions 
in $p+\bar{p}$ collisions at different $\sqrt{s}$ energy values.}
\label{fig:qcdweak}
\end{figure}

We study the sensitivity of the above results on the scale choice
$\kappa$, and find that varying $\kappa$ between 1/2 and 3/2 makes the
ratios vary a few ten percents (which is the usual sensitivity of
leading order QCD results on the factorization scale). The main
conclusion, namely the negligibility of weak interactions, remain the
same.

\section{Conclusion}
We have investigated the inclusive hadron production in $pp$ and 
$p\bar{p}$ collisions at FERMILAB TEVATRON and CERN LHC, including 
strong and weak-interaction channels. We have performed a LO calculation
to obtain an estimate on the hadron excess from weak channels.

Our parton model calculation confirms the usual practice of neglecting weak 
interaction contributions when calculating hadron spectra, in $pp$ and $p\bar{p}$ 
collisions at collision energies accessible with present technology. 
The weak-interaction contribution is always orders of magnitude smaller.

We expect that repeating our analysis at next-to-leading order accuracy
would lead to similar conclusion especially for the ratios 
$R_{p\bar{p}/pp}$ and $R_{W/S}$ where the effect of the NLO K factors
cancel to a large extent.

A naive estimate of comparing the running couplings at the momentum
scales involved, namely $(\alpha_{EW}/\alpha_s)^2 = O(10^{-1})$ is
still an overestimate of the ratio on Fig. 5, because of the dominance
of gluon channels in hadron production.

\acknowledgments
P.L. thanks Prof. B.L. Ioffe for stimulating discussion 
at the Gribov-80 Workshop. This work has been supported by
the OTKA Grant No. 77816 and 
the T\'AMOP 4.2.1./B-09/1/KONV-2010-0007 project.


\begin{thebibliography}{0}

\bibitem{ALICE}
  \Name{K. Aamodt et al., ALICE Collaboration}
  \REVIEW{Phys. Lett. B}{693}{2010}{53}.

\bibitem{ATLAS}
  \Name{ATLAS Collaboration}
  CERN-PH-EP-2010-079.

\bibitem{CMS}
  \Name{CMS Collaboration}
  CMS-PAS-QCD-10-008.

\bibitem{CDF}
  \Name{T. Altonen et al., CDF Collaboration}
  \REVIEW{Phys. Rev. D}{79}{2009}{112005}.

\newpage
\bibitem{CDFerr}
  \Name{T. Altonen et al., CDF Collaboration}
  \REVIEW{Phys. Rev. D}{82}{2010}{E119903}.

\bibitem{Albino}
  \Name{S. Albino, B.A. Kniehl \and G. Kramer}
  \REVIEW{Phys. Rev. Lett.}{104}{2010}{242001}.

\bibitem{Ioffe}
  \Name{B.L. Ioffe}
  arXiv:1005.1078.

\bibitem{FieldQCD}
  \Name{R.D. Field}
  \Book{Applications of Perturbative QCD}
  \Publ{Addison-Wesley Publishing, Reading}
  \Year{1989}.

\bibitem{MSTW}
  \Name{A.D. Martin, W.J. Stirling, R.S. Thorne \and G. Watt}
  \REVIEW{Eur. J. C}{63}{2009}{189}.

\bibitem{KKP}
  \Name{B.A. Kniehl, G. Kramer \and B. P\"otter}
  \REVIEW{Nucl. Phys. B}{582}{2000}{514}.

\bibitem{PDG}
  \Name{K. Nakamura et al., Particle Data Group}
  \REVIEW{J. Phys. G}{37}{2010}{075021}.

\bibitem{Szczurek2003}
  \Name{A. Szczurek}
  \REVIEW{Acta Phys. Polon. B}{35}{2004}{161}.


\end{thebibliography}
\end{document}